\documentclass[conference]{IEEEtran}
\IEEEoverridecommandlockouts
\usepackage{cite}
\usepackage{amsmath,amssymb,amsfonts}
\usepackage{algorithmic}
\usepackage{tikz}
\usepackage{pgfplots}
\usepackage{graphicx}
\usepackage{textcomp}
\usepackage{xcolor}
\usepackage{tabularx}
\usepackage{url}
\usepackage[flushleft]{threeparttable}
\usepackage{balance}
\usepackage{mathtools}

\usetikzlibrary{backgrounds}

\def\BibTeX{{\rm B\kern-.05em{\sc i\kern-.025em b}\kern-.08em
    T\kern-.1667em\lower.7ex\hbox{E}\kern-.125emX}}

\newcommand{\newlinecell}[2][c]{%
\begin{tabular}[#1]{@{}c@{}}#2\end{tabular}}

\allowdisplaybreaks

\begin{document}

\title{Efficient Pilot Allocation for URLLC Traffic\\in 5G Industrial IoT Networks
    \thanks{This work was supported by the National Science Centre, Poland,
    	under grant no. 2017/25/B/ST7/02313: ``Packet routing and transmission
    	scheduling optimization in multi-hop wireless networks with multicast
	traffic''. The work of Emma Fitzgerald was also partially supported by the
	Celtic-Next project 5G PERFECTA, the SSF
	project SEC4FACTORY under grant no. SSF RIT17-0032, and the strategic research
	area ELLIIT.} }

\author{{\setlength{\tabcolsep}{12pt}\begin{tabular}{ccc}\newlinecell{Emma Fitzgerald\IEEEauthorrefmark{1}\IEEEauthorrefmark{2}\\
    emma.fitzgerald@eit.lth.se}%
&\newlinecell{Micha{\l}
Pi\'{o}ro\IEEEauthorrefmark{2}\\
m.pioro@tele.pw.edu.pl}\end{tabular}}\\[1ex]
\begin{tabular}{cc}\newlinecell{\newlinecell{\IEEEauthorrefmark{1}\emph{Department of
Electrical and}\\\emph{Information Technology}}\\
\emph{Lund University}\\
Lund, Sweden}&%
\newlinecell{\IEEEauthorrefmark{2}\emph{Institute of Telecommunications}\\
\emph{Warsaw University of Technology}\\
Warsaw, Poland}\end{tabular}}

\maketitle
\vspace{-1.0cm}

\begin{abstract}

    In this paper we address the problem of resource allocation for alarm
    traffic in industrial Internet of Things networks using massive MIMO. We
    formulate the general problem of how to allocate pilot signals to alarm
    traffic such that delivery is guaranteed, while also minimising the number
    of pilots reserved for alarms, thus maximising the channel resources
    available for other traffic, such as industrial control traffic. We present
    an algorithm that fulfils these requirements, and evaluate its performance
    both analytically and through a simulation study. For realistic alarm
    traffic characteristics, on average our algorithm can deliver alarms within
    two time slots (of duration equal to the 5G transmission time interval)
    using fewer than 1.5 pilots per slot, and even in the worst case it uses around 3.5
    pilots in any given slot, with delivery guaranteed in an average of
    approximately four slots.

\end{abstract}


\section{Introduction} \label{sec:intro}

Ultra-reliable and Low Latency Communication (URLLC) is one of the target use
cases for 5G \cite{shafi20175g}. However, the performance requirements for this
type of traffic are not yet met for challenging use cases such as the industrial
Internet of Things (IoT). Moreover, little work has been done on the medium
access control (MAC) layer of massive multiple-input multiple-out (MIMO)
\cite{marzetta2016fundamentals}, a key technology for 5G, and mechanisms for
URLLC traffic are lacking.

In this paper, we present a MAC scheme for alarm traffic in industrial control
networks based on massive MIMO. In massive MIMO, pilots --- known signals needed
to obtain channel state information (CSI) for each user
\cite{marzetta2016fundamentals} --- are a limited resource. We formulate the
general problem of allocation of pilot signals to alarm sources, which trigger
alarms when unusual events occur, for example machine failures or control values
detected outside of a specified safe parameter range. We present an algorithm for
efficient pilot allocation that guarantees alarm delivery. We performed a
simulation study which shows good performance of our algorithm, with varying
numbers of alarms and alarm trigger probabilities. In the average
case, for an alarm trigger probability of $1$\% per alarm and slot, less than 1.5
pilots per time slot needed to be reserved for alarm traffic,
and alarms were delivered within two slots. In the worst case, alarms are delivered
within an average of just over $4$ slots, with at most $3.5$ pilots needed per
slot. The length of a time slot in our scheme is equal to the transmission time
interval, which in 5G can be as short as 125 $\mu$s \cite{sachs20185g}.

With the rise of Industry 4.0 comes a greater need for flexibility in
manufacturing and other industrial processes
\cite{vaidya2018industry,lu2017industry}.  Key features of Industry 4.0 include
optimisation and customisation of production, automation and adaption, and
automatic data exchange and communication, while real-time capability,
decentralisation, and modularity are some of the important operating principles
\cite{lu2017industry}. These trends will be facilitated by the transition to
wireless communications for industrial control processes, allowing for cheaper
and more scalable communications, with reduced cabling costs and the ability to
easily reconfigure the factory floor. However, as yet, the development and
deployment of wireless protocols suitable for real-time communication in this
context is limited, since existing protocols are not able to
meet the stringent requirements of industrial control traffic in terms of
latency and allowable packet loss rates
\cite{wollschlaeger2017future,gidlund2017will}.

Massive MIMO is a promising technology in this domain. Diversity gain can
greatly increase the reliability of communications for industrial systems
\cite{holfeld2016wireless}; for example, with no diversity gain, a $90$ dB
margin is needed in the fading gain in order to reach a channel outage
probability suitable for factory or process automation, while with a diversity
order of 15, this is reduced to $9$ dB. In massive MIMO, the use of up to
hundreds of antennas inherently provides a high degree of spatial diversity,
equal to the number of antennas when using maximum ratio combining, or the
number of antennas minus the number of concurrent users for zero forcing
\cite{marzetta2016fundamentals}. Moreover, an effect known as channel hardening
\cite{larsson2014massive,gunnarsson2018channel} makes massive MIMO channels
behave more like wired channels, smoothing out channel variations and providing
predictable performance. This effect is facilitated by low correlation between
the channels of different users. The otherwise challenging radio environment of
a factory \cite{holfeld2016wireless} is thus turned into a strength, as its
metallic fixtures and moving machine tools and robots provide a rich multipath
channel propagation environment that aids the differentiation of user signals in
massive MIMO systems.

In order to take advantage of the benefits promised by massive MIMO for
industrial communications, suitable MAC protocols are needed. However, thus
far, work on massive MIMO MAC for the IoT is limited, especially when it comes
to URLLC. A few recent works consider random access and/or grant-free protocols
for massive MIMO in IoT scenarios
\cite{bjornson2016random,sorensen2014massive,de2016random}. In particular,
\cite{liu2018sparse} argues for the advantages of massive MIMO over other
technologies in the presence of massively many devices, each only transmitting
intermittently. Scheduled access for periodic IoT traffic has also been
considered in \cite{fitzgerald2019massive}. However, none of these works cater
to the particular needs of industrial control traffic, in terms of packet loss
and latency guarantees, which we address in this paper.

The rest of this paper is organised as follows. In Section~\ref{sec:problem}, we
formulate the pilot allocation problem and define our traffic model and
performance requirements. Section~\ref{sec:algorithm} then presents our pilot
allocation algorithm based on a new concept that we call \emph{collision trees}.
Section~\ref{sec:simulation} details our simulation-based performance study and
results, and finally Section~\ref{sec:conclusion} concludes this paper.

\section{Pilot Allocation for Alarm Traffic}\label{sec:problem}

The problem we will address is pilot allocation for alarm traffic in a 5G
industrial Internet of Things scenario. Typically, in such a use case a
dedicated network (or network slice) would be used, and so we do not consider
traffic for other applications. In this scenario, a massive MIMO base station
provides communications for both continuous control traffic and sporadic, but
critical, alarm traffic. In order to communicate, both these traffic classes
must be allocated pilot signals, in such a way as to guarantee the delivery of
alarms, while also maintaining high performance for the control traffic. 

\subsection{Massive MIMO Transmission}\label{sec:MaMIMO_transmission}

Here we will provide a brief overview of transmission in massive MIMO systems; a
more comprehensive treatment can be found in \cite{marzetta2016fundamentals}.
Transmission in time division duplex-based massive MIMO occurs in
\emph{coherence blocks}, consisting of a time interval and frequency band across
which the channel is constant, to within a small margin of error. Each coherence
block contains a number of channel resource elements, which can each be used for
downlink or uplink data, or (uplink) pilot signals. Pilot signals are used to
measure channel state information for each user, and must be orthogonal. The CSI
is then fed into the preprocessing matrix used to direct each data stream to its
respective recipient on the downlink. or differentiate each incoming data stream
on the uplink.

If two users transmit the same pilot signal in the same coherence
block, the CSI for these users will be inaccurate and
data to or from them will be encoded or decoded incorrectly. This
phenomenon is known as \emph{pilot contamination}, and effectively results in a
collision between the user transmissions in the same way as interference in
single-antenna wireless systems does. Collisions can be avoided by allocating each
user a unique, orthogonal pilot, but for sporadic alarm traffic, this would result in
a very inefficient use of resources, since most of the time these pilots will go
unused.

One key difference between collisions due to pilot contamination and collisions
in single-antenna systems is that with the former, the base station is able to
communicate to all users involved in a collision via a multicast transmission
using their combined CSI; effectively, the contaminated pilot provides CSI for
all the involved users as a group. If the users in the group have very different
received power, some of them may fail to receive such a transmission, however,
in our scenario this is unlikely, since we have small, indoor cells and so no
user will be very far from the base station. Multicast transmissions to a
specific group of users involved in a collision give us new possibilities to
handle collisions, for example by allocating dedicated resources (pilots) for
collision resolution. In our proposed algorithm in Section~\ref{sec:algorithm},
we will make use of this capability.

\subsection{Traffic Classes and Requirements}

Efficient pilot allocation depends on the traffic to be served, so as to strike
the right balance between wastage of resources due to collisions and that due to
unused pilots. In our scenario, the traffic consists of two classes: control
traffic and alarm traffic. Control traffic encompasses the transmissions between
machines on the factory floor and their controllers. This traffic is regular, in
some cases even deterministic, and has stringent latency requirements in order
to arrive within a specified control loop period.  In this work, however, alarm
traffic will be the main focus, while control traffic will be regarded as base
load traffic, for which we will evaluate the performance impact of serving the
alarm traffic. Alarms are infrequent and unpredictable, but nonetheless must be
delivered reliably, making resource allocation for alarm traffic challenging.
Our goal is thus to provide delivery guarantees for alarm traffic, while
minimising the pilot resources required to do so.

There are two key performance requirements for industrial automation traffic,
latency and packet loss probability, and their values depend on the
specific industrial automation domain. The domains we will focus on are
process automation, with an update frequency of 10 to 1000 ms, and factory
automation, with an update frequency of 500 $\mu$s to 100 ms
\cite{gidlund2017will}. These two domains are of interest in this work because
their update frequencies are sufficiently fast to require pilot allocation
strategies that cater specifically to them, while also slow enough to be
feasible to realise using the 5G transmission time interval, which ranges from
125 $\mu$s to 1 ms \cite{sachs20185g}.

The packet loss rate required for our targeted domains is $1 \times 10^{-9}$
\cite{gidlund2017will}. This extremely low loss rate includes all potential
causes of packet loss, so we will aim for 100\% delivery guarantees in our pilot
allocation strategy; that is, no packet should be lost due to a failure in
resource allocation (after collision resolution). The packet loss rate and
latency requirements should thus be considered jointly, as they are inherently
tied together by the pilot allocation scheme. Packet loss can still occur for
other reasons, such as noise, but this is beyond the scope of the current work. 

\subsection{Alarm Traffic Model}

The performance of any given pilot allocation strategy depends on the
characteristics of the alarm traffic. We will adopt the following alarm traffic
model. We define a window consisting of $T$ time slots.  Each slot represents
one coherence interval, in which there may be multiple coherence blocks at
different frequencies. Within each slot, there are $P$ pilots available in
total, and each pilot can be assigned to one or more users for that slot. If two
or more users are assigned the same pilot in a given slot, a collision can occur
if both of them transmit during the slot. A pilot can also be unused (wasted) if
its assigned user(s) do not transmit in the slot.

We have a set $\mathcal A$ of alarm sources, each of which represents one type
of alarm that can arise during the window. The window should be limited in time,
as defined above, for two reasons. First, the set of alarms that may occur can
change over time, for example when the factory floor is reconfigured. Second,
the time scale at which alarms need to be served may be very different to that
at which they are reset. For example, an alarm may be triggered upon the
malfunction of a given machine, necessitating that the machine be shut down
quickly. However, the time to repair the machine may be much longer, and while
the machine is nonoperational, it is no longer possible for that alarm to be
triggered again. The window should therefore be an interval in time in which
the set of possible alarms is constant.

Each alarm $a \in \mathcal A$ has a probability $p(a)$ to be triggered during
each slot, but once triggered, an alarm cannot be triggered again within the
window. While in reality, alarm trigger probabilities could be correlated, we
begin by assuming independent alarms to simplify the analysis, and will consider
correlated alarms in future work. Each triggered alarm also has a deadline: a
number of slots within which it must be successfully received by the base
station. If a given alarm is not successful on its first transmission attempt,
for example if there was a collision, it will attempt retransmission up until
its deadline according to the pilot allocation scheme. We assume that all alarm
messages are short enough to be transmitted within a single slot.

\subsection{Pilot Allocation Problem}\label{sec:pilot_allocation_problem}

Our problem is then to define a pilot allocation scheme that
guarantees delivery of all alarms within their deadlines. A naive strategy could
be to simply assign one pilot to each alarm, in every slot. This would certainly
guarantee alarm delivery, but would be very inefficient since the probability of
any alarm being triggered at all is very low, let alone in a given slot.
Moreover, if there are many possible alarms, then there may not be enough pilots
in each slot to uniquely assign one to each alarm source. Our goal will
therefore be to minimise the number of pilots that need to be assigned to
alarms, while still guaranteeing delivery.

Formally, we can define a pilot allocation scheme as a finite sequence of
pilots defined for each alarm source. A given alarm source, when its alarm is
triggered, begins by transmitting the alarm message in the next slot using the
first pilot in its pilot sequence. If there is a collision, it attempts
retransmission in the following slot, using the second pilot in its sequence.
The alarm source will know that there is a collision because it will fail to receive
an acknowledgement from the base station during the downlink phase of the
coherence block in which it transmits. Upon further collisions the alarm source
proceeds along the sequence, one slot at a time, until it reaches the end: its
final retransmission attempt. A pilot sequence may also contain blank pilots,
indicating that the alarm source remains silent in the corresponding slot instead of
attempting retransmission.

When a collision occurs, the base station can optionally transmit a \emph{pilot
offset} to all alarm sources involved in the collision, using a multicast
transmission as described in Section \ref{sec:MaMIMO_transmission}. In this
case, the alarm sources add the pilot offset to the next value in their pilot
sequences. For example, if alarm $a\in \mathcal A$ has pilot $1$ as its next
allocated pilot, and alarm $b\in \mathcal A$ has pilot $2$ as its next pilot,
then if the base station transmits a pilot offset of $10$, alarm $a$ will
transmit using pilot $11$ in the next slot, and $b$ will transmit using pilot
$12$. The pilot offset thus allows for dynamic pilot allocation by the base
station, while still having fixed, pre-determined sequences for each alarm
source. As we will see in Section \ref{sec:algorithm}, this can simplify
collision resolution, since each collision can be resolved in its own pilot
range in parallel with any new alarm transmissions.

A sufficient condition to guarantee alarm delivery within the deadline is that
each pilot sequence should be no longer than the number of slots from an alarm
triggering until its deadline, and that the last pilot in each alarm's sequence
should be unique across the set of alarms. If an alarm source has a unique pilot
allocated to it in a given slot, then it will be able to transmit its alarm
message without any risk of a collision, thus guaranteeing delivery in that
slot. Since there is no reason to continue retransmission once collision-free
transmission is guaranteed, we specify that the unique pilot condition should be
met for the last pilot in each sequence. Then, if the length of the sequence is
no more than the number of slots until the deadline, the alarm will be sent
without collision in the worst case during the last slot before the deadline.

\section{Pilot Allocation Using Collision Trees}\label{sec:algorithm}

In this section we will present an efficient pilot allocation algorithm that can
guarantee alarm delivery. Our algorithm is based on a concept we call
\emph{collision trees}, a method for assigning pilot sequences to alarms such
that alarms with lower trigger probabilities have longer sequences and share
pilots with other alarms more often. Shared pilots entail a risk of collision,
if at least two of the alarms sharing the pilot are triggered at the same time,
and therefore it is beneficial for lower probability alarms to share pilots more
than higher probability alarms. The authors' Python implementation of the
collision tree algorithm \cite{collision_tree} is available for download and use
under the GNU General Public License.

\subsection{Algorithm Description}\label{sec:algorithm_description}

We begin by allocating a single, common pilot for all alarms in all slots. The
common pilot is used for the initial alarm transmissions of alarm sources
without requiring a prior grant for channel access. In the best case, all alarms
triggered in the window will arrive in different slots, and no further pilots
will need to be allocated for them. as they will each transmit without collision
on the common pilot.

\begin{figure}
    \begin{center}
	\includegraphics[width=\columnwidth]{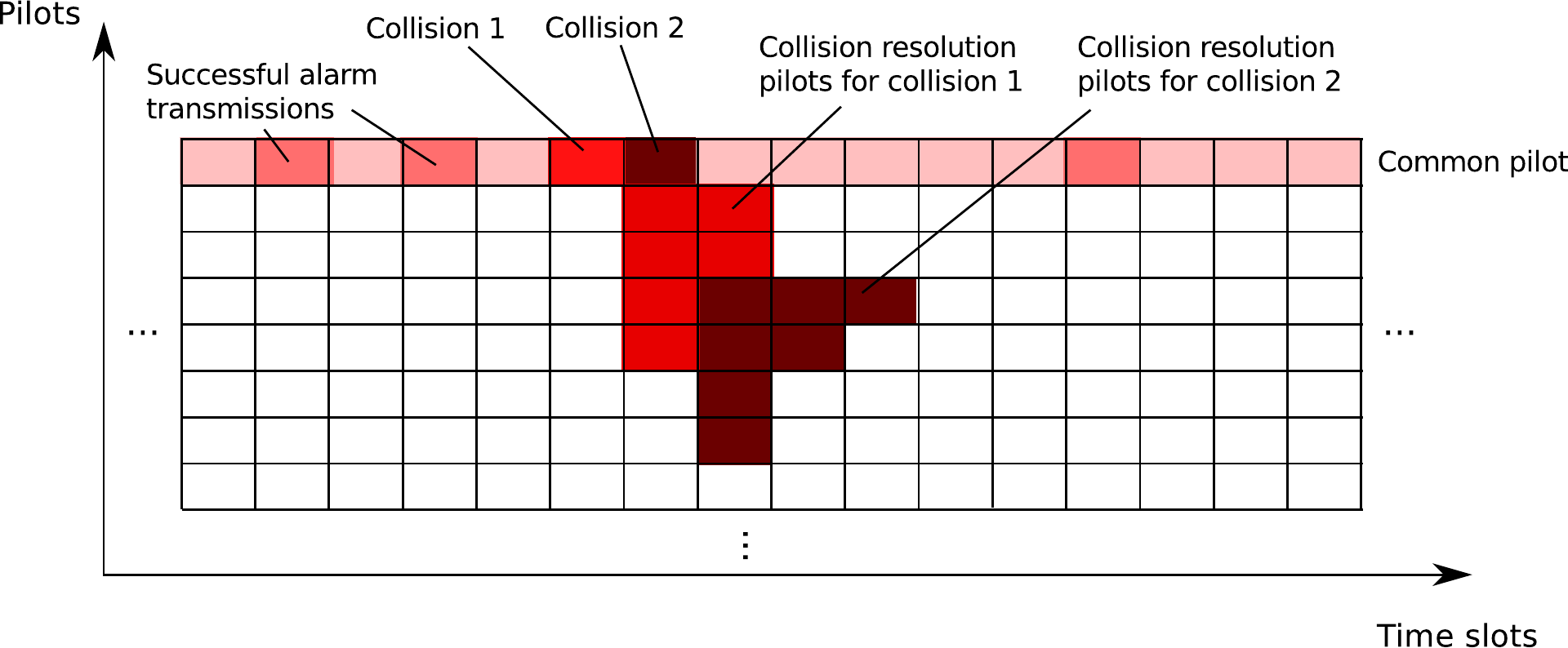}
	\caption{Example of pilot allocation for alarm transmission and
	    collision resolution. After collision 1, four pilots are allocated
	    in the following slot for collision resolution. This does not
	    resolve all alarms in the collision, so a further two pilots are
	    allocated in the next slot, after which the collision is resolved.
	    Similarly, Collision 2 requires first four pilots, then two, then
	    finally one in the third slot following the initial transmission.}
	\label{fig:pilot_allocation}
    \end{center}
\end{figure}

In the event of a collision, however, we resolve the collision in its own,
isolated pilot range, facilitated by the base station transmitting a pilot
offset to all involved alarm sources. This allows us to consider only the alarms
involved in the collision, and ignore any further alarms that arrive during
collision resolution, as these will be sent using the common pilot, and then, if
necessary, undergo collision resolution in a separate pilot range not affecting
the original collision resolution process. This somewhat reduces the efficiency
of collision resolution, however it greatly simplifies the design of the pilot
allocation scheme and allows us to more easily guarantee alarm delivery. An
example of pilot allocation with collision resolution is shown in
Figure~\ref{fig:pilot_allocation}.

Once a collision has occurred, the involved alarm sources attempt retransmission
according to their predefined pilot sequences (see Section
\ref{sec:pilot_allocation_problem}). To design these sequences, we will use
collision trees, inspired by the trees used to create Huffman codes
\cite{huffman1952method}. A collision tree $\mathcal T = (\mathcal V, \mathcal
E)$ is constructed as follows. We begin with the set of alarms $\mathcal A$,
together with their trigger probabilities $p(a)$, $a \in \mathcal A$. Each alarm
will become a leaf in the collision tree.  We then combine the two alarms with
the lowest probabilities, making them children of a parent node that also has a
probability associated with it, namely the probability of at least one of its
children (alarms) being triggered.

We then repeat the procedure, combining the two nodes with the lowest
probabilities that do not yet have parent nodes, by making them children of a
newly created parent node. The probability for the new node is given by the
probability of at least one of the alarms in its child subtree being triggered.
This process continues to iterate until finally we are left with a single node,
which will become the root node of the tree, and whose probability is equal to
the probability of any alarm being triggered in a given slot.

For a given node $v \in \mathcal V$ in the tree, the probability that at least
one of the alarms in its subtree is triggered is given by
\begin{equation}\label{eq:node_prob}
    \pi(v) = 1 - \prod_{l \in \mathcal L(v)} \left(1 - \pi(l)\right),
\end{equation}
where $\mathcal L(v)$ is the set of leaf nodes descended from $v$, and where $\pi(l)$,
for $l \in \mathcal V$ a leaf node of the tree, is equal to the probability of
the associated alarm being triggered. That is, $\pi(l) = p(a(l))$, where $a(l)
\in \mathcal A$ is the alarm associated with leaf node $l \in \mathcal V$.

Once the tree has been constructed, we need to assign pilots to each node in the
tree. Each level of the tree represents one time slot in the collision
resolution process, and so nodes in the tree must be assigned pilots unique
within their level. This is feasible so long as the number of nodes in any level
does not exceed the total number of pilots available in each time slot. To
assign pilots, we simply label the nodes in each level with pilot numbers,
starting from $1$ and increasing up to the number of nodes in the level. The
root of the tree is assigned the common pilot for initial alarm transmissions.
To determine the pilot sequence for a given alarm, we can then read, in order,
the pilots assigned to each node in the path from the root to the leaf node
corresponding to the alarm.

\begin{figure}
    \begin{center}
	\includegraphics[width=0.7\columnwidth]{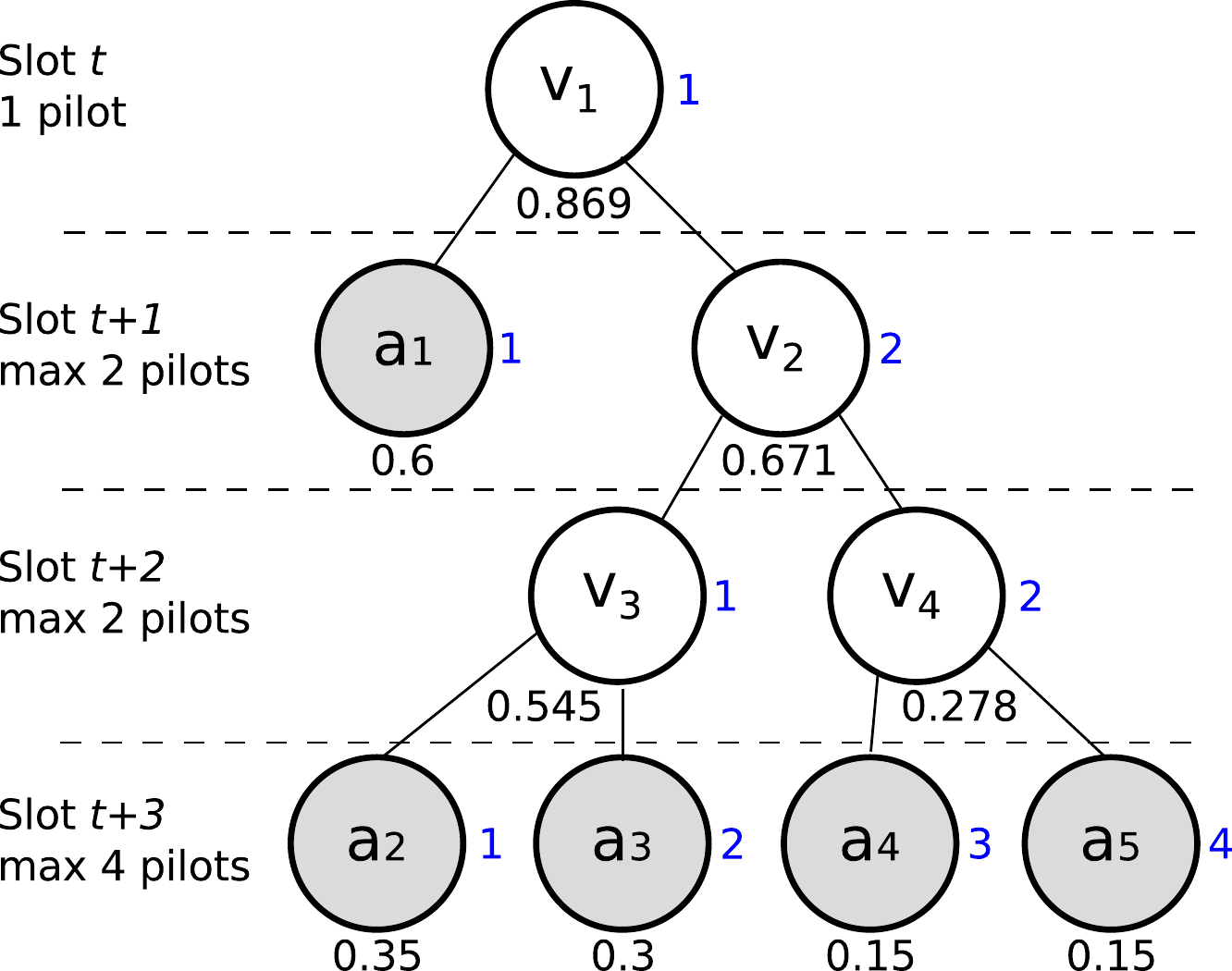}
	\caption{An example collision tree with five alarm sources. Leaf nodes
	    representing alarms are shown in grey. Each node's probability is
	    shown below it, and the pilot assigned to each node is shown in blue
	    to its right.}
	\label{fig:collision_tree}
    \end{center}
\end{figure}

An example of a collision tree is shown in Figure~\ref{fig:collision_tree}, for five
alarms $a_1\dots a_5$, with trigger probabilities $p(a_1) =0.6$, $p(a_2) =
0.35$, $p(a_3)=0.3$, $p(a_4)=0.15$, and $p(a_5)=0.15$. First, $a_4$ and $a_5$
are combined, since they have the lowest trigger probabilities, and placed as
children of a new node $v_4$. The probability for $v_4$ is given by $\pi(v_4) =
1 - (1 - 0.15)(1 - 0.15) = 0.278$. The next two orphan nodes (nodes without a
parent) with the lowest probabilities are then $a_2$ and $a_3$, which are placed
as the children of a new node $v_3$, with probability $1 - (1 - 0.3)(1 - 0.35) =
0.545$. Now the two orphan nodes with the lowest probabilities are $v_3$ and
$v_4$, so they are placed as the children of a new node $v_2$, with probability
$1 - (1 - 0.35)(1 - 0.3)(1 - 0.15)(1 - 0.15) = 0.671$. Finally, we only have two
orphan nodes remaining, so they are placed as the children of the new root node
$v_1$, with probability $0.869$, which is the probability of at least one of the
alarms triggering. The resulting collision tree has four levels, corresponding
to a maximum of four slots needed to resolve all alarms. The maximum number of
pilots that would need to be reserved in any slot is four in slot $t+3$, where
$t$ is the slot in which the alarms were first transmitted.

\subsection{Performance Analysis}\label{sec:analysis}

There are two key performance metrics that are of interest in assessing the
quality of a given pilot allocation scheme. The first is the delivery time for
the alarms, and the second is the expected number of pilots reserved for alarms,
as these then cannot be used for control traffic. The delivery time can be
considered both for each individual alarm, and for the entire set of alarms, by
taking aggregate measures across the set. The maximum delivery time for any
given alarm is given simply by the length of its pilot sequence. For collision
tree-based allocation, this is equivalently the length of the path from the leaf
node associated with the alarm to the root of the tree. We can express this as
$\hat D(a) = |\mathcal R(l(a))|$, where $\hat D(a)$ is the maximum delivery time
for alarm $a \in \mathcal A$ in time slots, $l(a) \in \mathcal V$ is the leaf
node in the tree associated with alarm $a$, and $\mathcal R(v)$, $v \in \mathcal
V$, is the path from a node $v$ to the root of the tree, consisting of the
parent node of $v$, followed by the parent node of $v$'s parent, and so on up to
the root node. The maximum delivery time can be used to determine whether or not
alarm $a$ can be guaranteed delivery within its deadline; if $\hat D(a)$ is less
than or equal to the number of time slots until $a$'s deadline, then delivery
within the deadline is ensured.

By default, our collision tree algorithm guarantees delivery of all alarms, but
not necessarily within their deadlines. However, by checking the maximum
delivery time of each alarm against its deadline, the tree can easily be
modified to ensure all alarms meet their deadlines, albeit with some loss of
performance. This can be done by moving an alarm node $l(a)$ up the tree along
$\mathcal R(l(a))$ until $\hat D(a)$ is within the deadline. First $l(a)$ is
moved up one level to become a sibling of its parent, with its grandparent node
as its new parent.  If the deadline is still not met, this procedure is repeated
so that $l(a)$ next becomes a sibling of its grandparent, then its
great-grandparent, and so on as needed.

In addition to the maximum delivery time, the expected delivery time for each
alarm is also important. Even though the deadline represents the absolute latest
time an alarm can be delivered, not all delivery times within the deadline are
necessarily equal: it may be beneficial to deliver the alarm sooner rather than
later. Our collision tree algorithm seeks to minimise the probability that an
alarm will use its entire pilot sequence in order to be delivered, and so the
expected delivery time will be significantly shorter than the maximum. Note that
while delivery time is analogous to codeword length in Huffman codes, its
analysis is more complicated because alarms do not always use their full pilot
sequence, that is, the expected delivery time for each alarm is in general not
equal to its maximum delivery time. Since Huffman codes are optimal in terms of
encoded message length \cite{huffman1952method}, our collision trees will also
be optimal in terms of maximum delivery times of alarms, but not necessarily in
terms of expected delivery time: more analysis is needed to prove or disprove
this result, and will be the subject of our future work.

The expected delivery time for an alarm $a \in \mathcal A$ depends on the
collision probabilities at each node $r \in \mathcal R(l(a))$ --- the
probability of a collision occurring on the pilot assigned to node $r$ during
its slot. This probability is given by
\begin{equation}\label{eq:coll_prob}
    c(v) = 1 - \smashoperator{\sum_{l \in \mathcal L(v)}} \pi(l)\,\, \smashoperator{\prod_{m \in \mathcal L(v)
    \setminus \{l\}}} \left( 1 - \pi(m) \right) - \smashoperator{\prod_{l \in
    \mathcal L(v)}} \left(1 - \pi(l) \right),
\end{equation}
that is, the complement of the probability that either none of the alarms
sharing the pilot assigned to $v$ were triggered, or only one of them was: as
soon as two or more alarms transmit, a collision will occur.

However, when determining the expected delivery time of a given alarm $a$, that
specific alarm must have been triggered, and so we need to find the conditional
collision probability given that $a$ was triggered. In this case, it is
sufficient for any one of the other alarms also covered by $v$ to be triggered,
and so the conditional probability of a collision at $v$ given $a$ was triggered is
\begin{equation}\label{eq:coll_prob_cond}
    c(v, a) = 1 - \prod_{m \in \mathcal L(v) \setminus \{ l(a) \}} \left( 1 -
    \pi(m) \right).
\end{equation}
The expected delivery time for alarm $a$ is then given by
\begin{equation}\label{eq:exp_delivery_time}
    E\left[ D(a) \right] = 1 + \sum_{r \in \mathcal R (l(a)) \setminus \{ v_0 \}} c(r, a).
\end{equation}
where $D(a)$ is the random variable representing the delivery time of alarm $a$,
and $v_0$ is the root node of the tree. The minimum delivery time is always one
slot, representing the case when the alarm is delivered straight away without
collision. In the event of a collision, an additional slot is then added for
each node along the path $\mathcal R(l(a)$ at which there is a collision. Note
that a collision at node $v \in \mathcal V$ also implies a collision at all
nodes along $\mathcal R(v)$, and the collision probability at any leaf node is
$0$. Finally, in order to obtain an aggregate performance metric across all
alarms, we can take the average delivery time across the alarms as
\begin{equation}\label{eq:exp_delivery_time_all}
    D = \frac{1}{|\mathcal A|} \sum_{a \in
    \mathcal A} E\left[ D(a) \right].
\end{equation}

Our second performance consideration is the effect of our pilot allocation
scheme on the control traffic, in terms of the total expected number of pilots
reserved for alarm traffic per slot, including pilots allocated for collision
resolution. However, calculating the expected number of reserved pilots with our
alarm traffic model is not straightforward, since once an alarm is triggered, it
can no longer be triggered again. It is thus removed from the set of possible
alarms, and consequently, the probability of a collision will be lower in
subsequent slots, meaning that the expected number of reserved pilots for each
slot will reduce monotonically as we move through the time window and alarms are
triggered. To calculate the expectation over all slots, we would therefore need
to consider all possible combinations of which alarms could trigger in the same
slots, and in which order, which is not feasible to do in practice.

We will therefore take as a performance measure the a priori expected number of
pilots reserved per slot, that is, the expected number of pilots needed for
alarms per slot assuming all alarms can be triggered. This will give an upper
bound on the true expectation where alarms can only be triggered once. The a
priori expected number of pilots reserved for alarm traffic is given by
\begin{equation}\label{eq:exp_num_pilots}
    \hat E[P] = 1 + \sum_{v \in \mathcal V \setminus \{v_0\}} c(R(v)),
\end{equation}
where $\hat E$ denotes the a priori expectation as defined above, $P$ is the
random variable representing the number of pilots reserved for alarm traffic per
slot, and $R(v)$, $v \in \mathcal V$, is the parent node of $v$. One pilot is
needed for each node whose parent node has a collision, with an additional pilot
for the root node (the common pilot reserved in all blocks for alarms).

The two performance metrics discussed here do not take into account the effect
of alarms being removed from the set of possible alarms after they have been
triggered, and so will overestimate the real performance. However, they provide
an indication of the worst case performance, when no alarms have yet been
triggered, and so assist with dimensioning the network. Further, these
metrics provide a means to compare different collision trees, produced either by
the algorithm given in Section~\ref{sec:algorithm} or others. For example, these
metrics could be used to optimise the collision tree, which we intend to explore
in our future work.

\section{Simulation Study}\label{sec:simulation}

In order to empirically test the performance of the collision tree algorithm, we
implemented a simulator in Python for alarm pilot allocation, along with
numerical functions for the performance metrics detailed in
Section~\ref{sec:analysis}. The full code is available online
\cite{collision_tree}.

\subsection{Simulation}

The simulation proceeds as follows. First, a set of alarms $\mathcal A$ is
generated, each $a \in \mathcal A$ with a per time slot trigger probability
$p(a)$ between $0$ and $p$, where $p$ is a simulation parameter. We
tested the following values of $p$: $0.001$, $0.005$, $0.01$, $0.05$, $0.1$, and
$0.5$. While some of these trigger probabilities are unrealistically high for
practical scenarios, they are useful to find the limits of the system's
capabilities. The number of alarms in $\mathcal A$, that is $|\mathcal A|$, was also
varied, from $10$ to $100$ in steps of $10$. Note that this is the total number
of possible alarms that can arrive, but the actual triggering of these alarms
depends on the trigger probabilities and not all will necessarily be triggered
during any given simulation.

For each configuration of these two parameters, $20$ problem
instances (instances of $\mathcal A$ along with $p(a) \in [0, p]$ for each $a \in
\mathcal A$) were generated. For each such instance, a collision tree was built
from the trigger probabilities $p(a)$ using the algorithm described in
Section~\ref{sec:algorithm_description}. From the collision tree, we calculated the analytical
performance metrics, specifically, the average alarm delivery
time (Equation~\eqref{eq:exp_delivery_time_all}) and the a priori expected
number of pilots used per slot (Equation~\eqref{eq:exp_num_pilots}). Finally,
the actual simulation was then run $50$ times for each instance, with a window
size of $50$ slots.

During the simulation, in each slot, a random number is drawn
between $0$ and $1$ for each alarm that has not yet been triggered. If this
number is lower than the alarm's trigger probability $p(a)$, the alarm is triggered in
that slot. If more than one alarm is triggered, a collision occurs and is
resolved using the collision tree in subsequent slots. Once the window has ended,
no more alarms are triggered, but the simulation continues until all remaining
collisions are resolved.

\subsection{Results}

\begin{figure}
    \begin{center}
	\includegraphics[width=0.8\columnwidth]{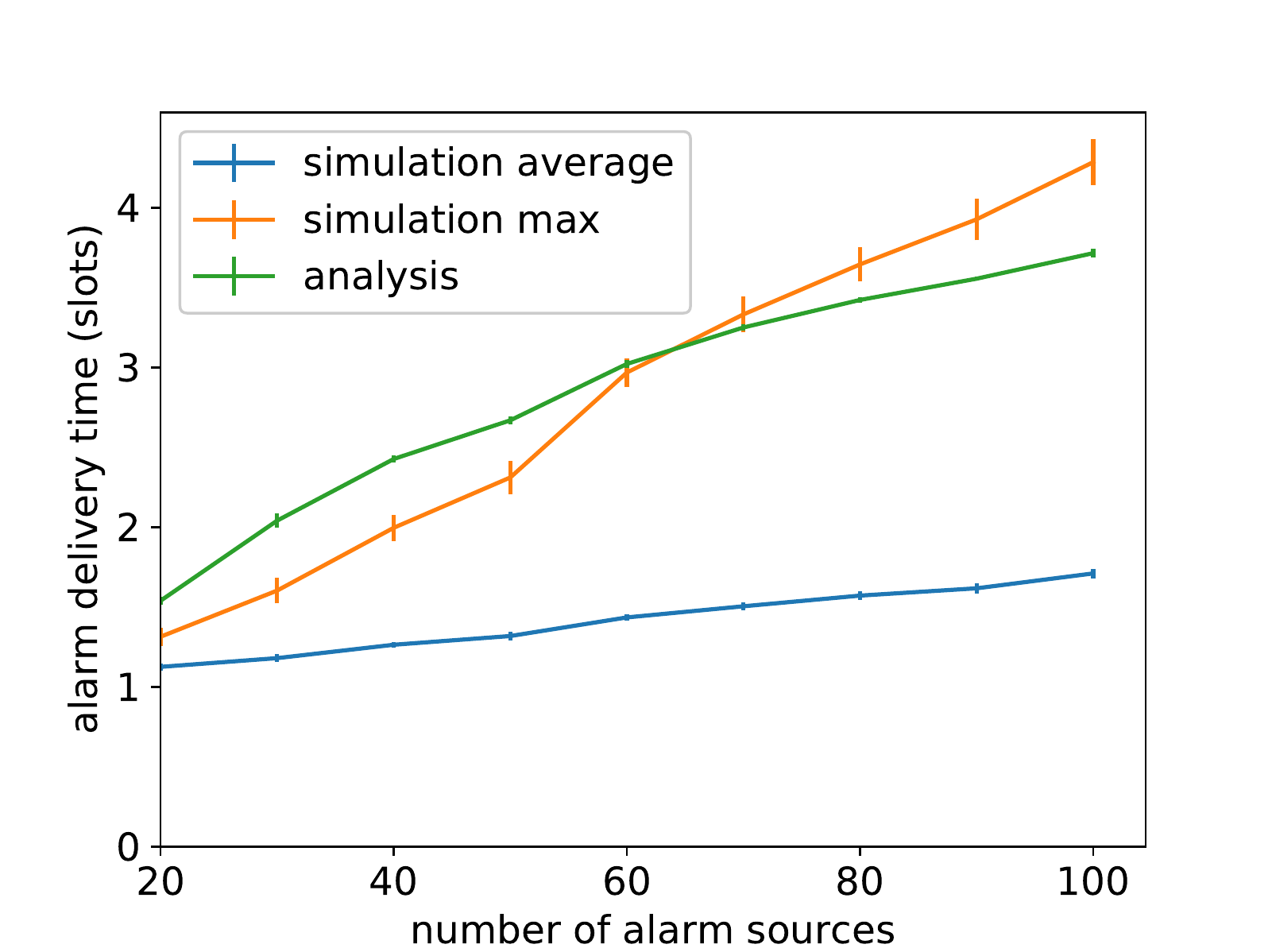}
	\caption{Delivery times for alarms with trigger probability upper bound $p =
	    0.01$, average across all simulation runs and instances (Simulation
	    average), maximum for each simulation run, averaged across all runs and
	    instances (Simulation max), and analytical average delivery time $D$
	    (Analysis, Equation~\eqref{eq:exp_delivery_time_all}).}
	\label{fig:delivery_time_0.01}
    \end{center}
\end{figure}

\begin{figure}
    \begin{center}
	\includegraphics[width=0.8\columnwidth]{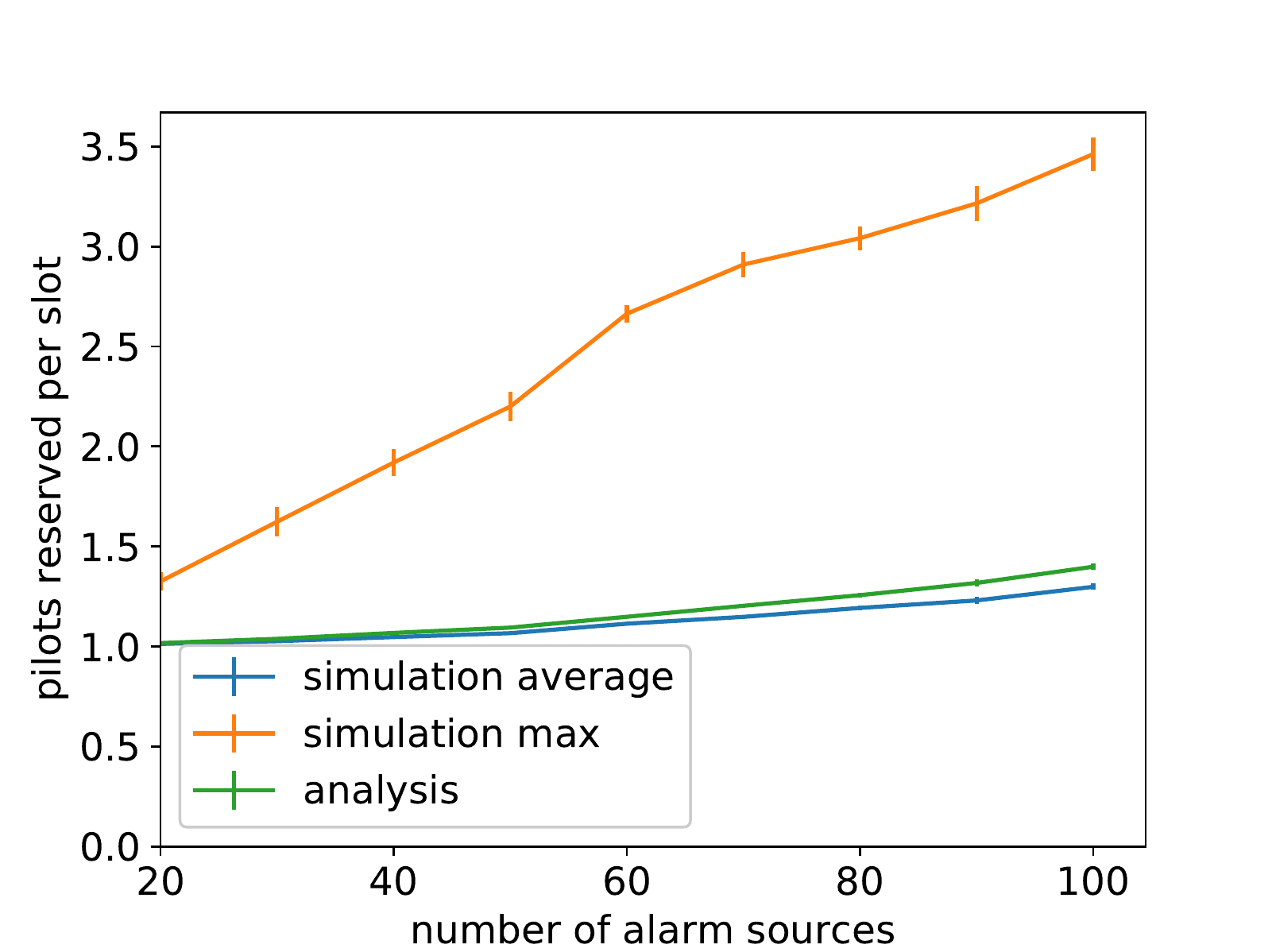}
	\caption{Pilots reserved for alarms per slot with trigger probability
	    upper bound $p = 0.01$, average across all simulation runs and
	    instances (Simulation average), maximum for each simulation run,
	    averaged across all runs and instances (Simulation max), and
	    analytical average delivery time $\hat E[P] $ (Analysis,
	    Equation~\eqref{eq:exp_num_pilots}).}
	\label{fig:pilots_per_slot_0.01}
    \end{center}
\end{figure}

A selection of simulation results are shown in
Figures~\ref{fig:delivery_time_0.01} and \ref{fig:pilots_per_slot_0.01}. Full
results can be obtained by downloading and running the simulation code from
\cite{collision_tree}. We show here the results for $p=0.01$, which gives
relatively high but nonetheless realistic trigger probabilities. For some
simulations, no alarms are triggered. In these cases, both the average and
maximum delivery times were set to $1.0$, as were the average and maximum pilots
per slot, since an alarm always needs at least one pilot in one slot to be
delivered and this thus gives a minimum value. All results are shown with $95\%$
confidence intervals,

The maximum delivery time for a given alarm can be theoretically as high as
$|\mathcal A|$ slots in the worst case, when the collision tree is unbalanced
such that there is one leaf for each level in the tree. However, as the delivery
time results (Figure~\ref{fig:delivery_time_0.01}) show, in practice the
delivery times are much shorter using our collision tree algorithm, on average
less than 2 slots and with a maximum of around $4$ slots for $p=0.01$.
The analytical average delivery time tracked the maximum, giving an
indication of the worst case performance, and it provides an upper bound for
the actual average, since it does not take into account alarms being removed
after they are triggered. Delivery times increased with $p$,
however even for the highest value tested, $p=0.5$, and with $100$ alarms, the
maximum delivery time was less than $8$ slots and the average around $4$ slots,
showing that our algorithm effectively controls the delivery time even in
challenging cases.

The collision tree algorithm also showed good performance with regards to the
number of pilots reserved for alarms per slot
(Figure~\ref{fig:pilots_per_slot_0.01}). Here, the analytical metric $\hat E[P]$
followed the simulation average for low to medium values of $p$, although at
higher values the analysis diverges from the simulation average. The
maximum number of pilots reserved for alarms was approximately $3.5$ pilots with
$100$ alarms and a maximum trigger probability of $0.01$. This shows that the
disruption to control traffic is limited, even in cases where many alarms are
triggered at once. Moreover, the average number of pilots reserved per slot was
very low, less than $1.5$. The collision tree algorithm is thus able to
guarantee alarm delivery while making very efficient use of pilot resources for
realistic alarm traffic. As $p$ increases, both the average and maximum number
of pilots used also increase, with a maximum number of pilots of $17.5$ for
$p=0.1$ and $56$ for $p=0.5$. This gives an indication of the maximum traffic
that can be accommodated by a single base station using our approach.

\section{Conclusion}\label{sec:conclusion}

\balance
In this paper we have studied a new problem in URLLC traffic in 5G networks,
that of massive MIMO pilot allocation for alarm traffic in industrial Internet
of Things scenarios, in particular factory and process automation. We have
presented a grant-free random access scheme for alarm traffic, together with an
algorithm for pilot collision resolution that can guarantee alarm delivery while
making efficient use of pilot resources.  In our future work we plan to further
investigate the performance when alarm deadlines are shorter than their initial
assigned pilot sequence length, necessitating moving some alarms further up the
collision tree. We also aim to find optimal collision trees and compare them
with those generated using our algorithm, as well as compare the performance of
collision trees with other contention resolution methods. Other possible
directions for future work include modelling and simulation of control traffic,
as well as studying different traffic distributions for alarms and correlation
between alarm sources.

\bibliographystyle{IEEEtran}
\bibliography{industry4.0,industrial_5G,massive_MIMO,references}

\end{document}